\documentclass[english,notitlepage,reprint,prl,superscriptaddress]{revtex4-1}
\usepackage[T1]{fontenc}
\usepackage[latin9]{inputenc}
\usepackage{color}
\usepackage{babel}
\usepackage{textcomp}
\usepackage{amssymb}
\usepackage{graphicx}
\usepackage[unicode=true,pdfusetitle,
 bookmarks=true,bookmarksnumbered=false,bookmarksopen=false,
 breaklinks=false,pdfborder={0 0 1},backref=false,colorlinks=true]
 {hyperref}

\makeatletter
 
 \@ifundefined{textcolor}{}
 {%
   \definecolor{BLACK}{gray}{0}
   \definecolor{WHITE}{gray}{1}
   \definecolor{RED}{rgb}{1,0,0}
   \definecolor{GREEN}{rgb}{0,1,0}
   \definecolor{BLUE}{rgb}{0,0,1}
   \definecolor{CYAN}{cmyk}{1,0,0,0}
   \definecolor{MAGENTA}{cmyk}{0,1,0,0}
   \definecolor{YELLOW}{cmyk}{0,0,1,0}
 }

\makeatother

\begin{document}

\title{Local and bulk $^{13}$C hyperpolarization in NV-centered diamonds
at variable fields and orientations}

\author{Gonzalo A. \'Alvarez}

\thanks{These authors contributed equally to this work.}

\affiliation{Department of Chemical Physics, Weizmann Institute of Science, Rehovot,
76100, Israel}

\author{Christian O. Bretschneider}

\thanks{These authors contributed equally to this work.}

\affiliation{Department of Chemical Physics, Weizmann Institute of Science, Rehovot,
76100, Israel}

\author{Ran Fischer}

\thanks{These authors contributed equally to this work.}

\affiliation{Department of Physics, Technion, Israel Institute of Technology,
Haifa, 32000, Israel}

\author{Paz London}

\affiliation{Department of Physics, Technion, Israel Institute of Technology,
Haifa, 32000, Israel}

\author{Hisao Kanda}

\affiliation{National Institute for Materials Science, 1-1 Namiki, Tsukuba, Ibaraki
305-0044, Japan}

\author{Shinobu Onoda}

\affiliation{Japan Atomic Energy Agency, 1233 Watanuki, Takasaki, Gunma 370-1292,
Japan}

\author{Junichi Isoya}

\affiliation{Research Center for Knowledge Communities, University of Tsukuba,
1- 2 Kasuga, Tsukuba, Ibaraki 305-8550, Japan}

\author{David Gershoni}

\affiliation{Department of Physics, Technion, Israel Institute of Technology,
Haifa, 32000, Israel}

\author{Lucio Frydman}

\email{Lucio.Frydman@weizmann.ac.il}

\selectlanguage{english}%

\affiliation{Department of Chemical Physics, Weizmann Institute of Science, Rehovot,
76100, Israel}

\begin{abstract}
Polarizing nuclear spins is of fundamental importance in biology,
chemistry and physics. Methods for hyperpolarizing $^{13}$C nuclei
from free electrons in bulk, usually demand operation at cryogenic
temperatures. Room-temperature approaches targeting diamonds with
nitrogen-vacancy (NV) centers could alleviate this need, but hitherto
proposed strategies lack generality as they demand stringent conditions
on the strength and/or alignment of the magnetic field. We report
here an approach for achieving efficient electron\textrightarrow $^{13}$C
spin alignment transfers, compatible with a broad range of magnetic
field strengths and field orientations with respect to the diamond
crystal. This versatility results from combining coherent microwave-
and incoherent laser-induced transitions between selected energy states
of the coupled electron-nuclear spin manifold. $^{13}$C-detected
Nuclear Magnetic Resonance (NMR) experiments demonstrate that this
hyperpolarization can be transferred via first-shell or via distant
$^{13}$Cs, throughout the nuclear bulk ensemble. This method opens
new perspectives for applications of diamond NV centers in NMR, and
in quantum information processing.
\end{abstract}
\maketitle
Nuclear spins are the central actors in NMR and magnetic resonance
imaging (MRI) \cite{Abragam1961a,Callaghan1993}. They are also promising
vehicles for storing and manipulating quantum information, thanks
to their long relaxation times \cite{gershenfeld_bulk_1997,Kane1998}.
At room temperature, however, nuclear polarization is very weak, resulting
in severe limitations in these spins' applicability. Dynamic nuclear
polarization (DNP) \cite{Abragam1978,Ardenkj??r-Larsen2003,Griesinger2012}
can bypass these limitations by transferring spin polarization from
electrons to nuclei, yet for this process to be efficient cryogenic
temperature operations are usually required \cite{Golman2006,Joo2006,King2010}.
Electronic spins of NV centers are promising alternatives for polarizing
nuclear spins in single crystal diamonds at room temperature conditions
\cite{Dutt2007,Jacques2009,Fischer2013,Wang2013}. Hitherto proposed
electron\textbf{\textrightarrow }nuclear polarization transfer methods,
however, have so far demanded finely tuned energy matching conditions,
such as the field strength and its orientation with respect to the
diamond crystal, in order to perform efficiently. These prerequisites
pose an obstacle for utilizing these methods in generic applications,
including the use of diamond powders in MRI or NMR-based analyses.
It is hereby shown that these demands for achieving a robust electron\textbf{\textrightarrow }nuclear
polarization transfer can be relaxed by the inclusion of incoherent
processes \cite{Erez2008,Alvarez2010}. The method hereby proposed
includes a continuous microwave (MW) irradiation that coherently addresses
the $0\leftrightarrow-1$ (or +1) $S=1$ electronic spin transition,
while exploiting the asymmetry of the electron-nucleus hyperfine interaction
\cite{Felton2009,Smeltzer2011,Dreau2012,Shim2013}. This selective
MW addressing is combined with the incoherent spin repopulation effects
introduced by the optical pumping process \cite{Gruber1997,Jelezko2004},
to produce an imbalance between the populations of the coupled nuclear
spins. This achievement of steady-state $^{13}$C polarization is
here demonstrated over a range of magnetic field strengths and orientations
as well as of electron-nuclear hyperfine interactions, both in optical
studies of single NV-centers and in shuttled NMR measurements of bulk
samples. Moreover, by exploiting the versatility of this new approach,
we shed light on the minor roles played by hyperfine-imposed spin-diffusion
barriers in the achievement of bulk nuclear spin hyperpolarization.

\section*{Results}

\begin{figure}
\includegraphics[width=1\columnwidth]{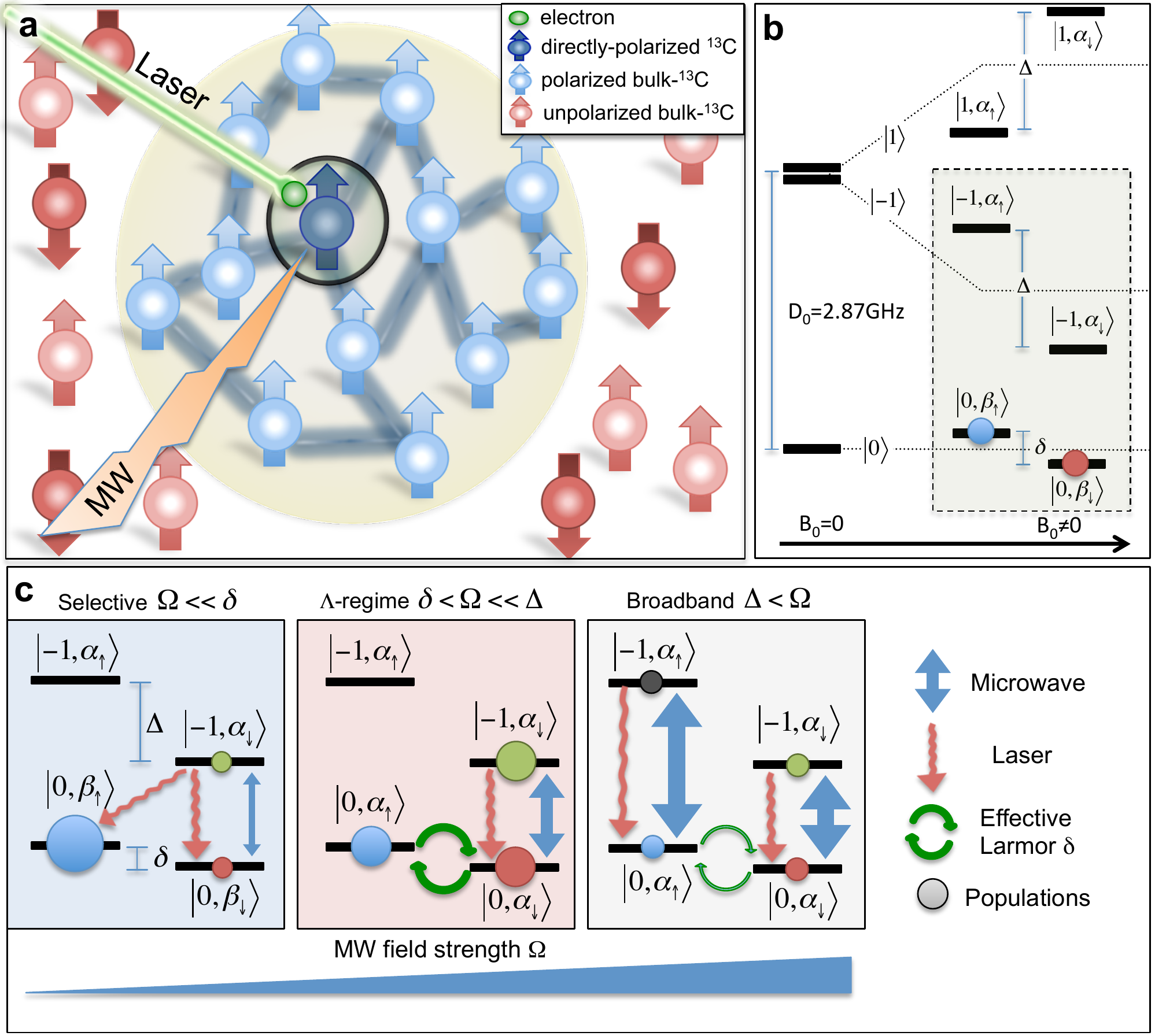}

\protect\caption{\label{fig:model_scheme}\textbf{MW-driven $^{13}$C polarization
derived from optically-pumped NV centers}. (\textbf{a}) Spin-1 NV
electronic defect strongly coupled to a $^{13}$C nucleus (black circle),
irradiated simultaneously by microwave and laser fields. The optically
pumped NV-spin transfers its polarization to the coupled nucleus,
and eventually to the remaining $^{13}$C (the bulk) by interactions
within a dipolar spin-network. (\textbf{b}) Energy level diagram of
the electron defect, HF-coupled to a $^{13}$C in presence of a potential
magnetic field $B_{0}$ (without MW fields). The dashed box shows
the energy levels addressed, and stresses an initial state containing
equal populations on the lower $\left|0,\beta_{\downarrow}\right\rangle $,
$\left|0,\beta_{\uparrow}\right\rangle $ eigenstates after optical
pumping. (\textbf{c}) Spin dynamical regimes determined by the relation
between the MW power and the energy splittings. The solid black lines
only represent the eigenstates $\left|0,\beta_{\downarrow}\right\rangle $,
$\left|0,\beta_{\uparrow}\right\rangle $, $\left|-1,\alpha_{\downarrow}\right\rangle $
and $\left|-1,\alpha_{\uparrow}\right\rangle $ of the system in the
``selective regime''; in the remaining cases these lines represent
states $\left|0,\alpha_{\downarrow}\right\rangle $, $\left|0,\alpha_{\uparrow}\right\rangle $,
$\left|-1,\alpha_{\downarrow}\right\rangle $ and $\left|-1,\alpha_{\uparrow}\right\rangle $
that are relevant for the MW selection rules, but where $\left|0,\alpha_{\downarrow}\right\rangle $
and $\left|0,\alpha_{\uparrow}\right\rangle $ are linear superpositions
of the eigenstates $\left|0,\beta_{\downarrow}\right\rangle $ and
$\left|0,\beta_{\uparrow}\right\rangle $. Blue vertical arrows represent
the MW excitation, circular green arrows represent an effective Larmor
precession with frequency $\delta$, while curly red arrows represent
a laser-induced relaxation-like process conserving the nuclear spin
state but driving the incoherent $\left|-1,\alpha_{\downarrow}\right\rangle \rightarrow\left|0,\alpha_{\downarrow}\right\rangle $
optical pumping \cite{Fuchs2012}. The filled and colored circles
schematize the populations of each state resulting from these dynamics.\textbf{}}
\end{figure}

\textbf{The MW-driven, relaxation-aided polarization transfer scheme.
}Figure \ref{fig:model_scheme}a defines the relevant system and active
interactions that will be used to introduce our polarization transfer
proposal. The system involves a single NV $S=1$ spin exhibiting a
ground-state zero-field splitting $D_{0}=2.87$ GHz. This electronic
spin is coupled to nearby or distant $^{13}$C nuclear spins ($I=1/2$)
through hyperfine (HF) interactions, and can be optically pumped to
populate the electronic $m_{s}=0$ level \cite{Gruber1997,Jelezko2004}.
The application of a weak static magnetic field ($\gamma_{e}B_{0}\ll D_{0}$),
induces an additional splitting of the electronic/nuclear states,
leading to the energy level diagram in Fig. \ref{fig:model_scheme}b.
Our scenario assumes a selective MW irradiation that solely addresses
the electron $\left|0\right\rangle \leftrightarrow\left|-1\right\rangle $
spin transitions, allowing one to treat the system as a four-level
manifold (Fig. \ref{fig:model_scheme}b). The nuclear spin components
of the eigenstates associated with $m_{s}=0$ ($\left|\beta_{\uparrow}\right\rangle $
and $\left|\beta_{\downarrow}\right\rangle $) and $m_{s}=-1$ ($\left|\alpha_{\uparrow}\right\rangle $
and $\left|\alpha_{\downarrow}\right\rangle $) have different quantization
axes, due to the asymmetry imposed by a hyperfine interaction that
is absent if the electron state is $m_{s}=0$ and present if $m_{s}=-1$
\cite{Felton2009,Smeltzer2011,Dreau2012,Shim2013}. The corresponding
eigenenergies exhibit splittings $\delta$ and $\Delta$ for the 0
and -1 manifolds, respectively. These differences in level splitting
and in quantization axes among nuclear spins that are initially unpolarized,
are here exploited to hyperpolarize them. Although not explicitly
shown in these energy level diagrams, the electronic and nuclear spins
involved in these manifolds are also coupled by dipole-dipole interactions
to the $^{13}$C ensemble via a spin-coupling network, enabling further
polarization transfers to the bulk (Fig. \ref{fig:model_scheme}a).

Different polarization transfer routes can be activated by irradiating
this HF-coupled system, that we discriminate depending on the relative
MW field strength $\Omega$ vis-a-vis the energy splittings $\delta$
and $\Delta$. In the selective $\Omega\ll\delta,\Delta$ regime (Fig.
\ref{fig:model_scheme}c), the MW-induced transitions involve solely
two directly-addressed eigenstates. Therefore the population of a
third, non-addressed state associated with the $m_{S}=0$ manifold,
grows systematically as driven by the laser-induced relaxation processes.
This generates an imbalance between the $\left|0,\beta_{\downarrow}\right\rangle $
and $\left|0,\beta_{\uparrow}\right\rangle $ populations, equivalent
to a nuclear polarization whose direction is defined by $m_{s}=0$.
As the MW power increases a ``$\Lambda-$regime'' where transitions
are induced among three eigenstates \cite{Shim2013}, is reached (Fig.
\ref{fig:model_scheme}c). We describe this $\delta\lesssim\Omega\ll\Delta$
regime in a basis set $\left\{ \left|0,\alpha_{\uparrow}\right\rangle ,\left|0,\alpha_{\downarrow}\right\rangle ,\left|-1,\alpha_{\downarrow}\right\rangle \right\} $;
here $\left|0,\alpha_{\uparrow}\right\rangle $ is a ``dark state''
for the MW in the sense that $\left\langle 0,\alpha_{\uparrow}\right|S_{x}\left|-1,\alpha_{\downarrow}\right\rangle =0$,
and $\left|0,\alpha_{\downarrow}\right\rangle $ is a ``bright state''
$\left\langle 0,\alpha_{\downarrow}\right|S_{x}\left|-1,\alpha_{\downarrow}\right\rangle =1$
that is addressed by the MW. Due to the different HF properties associated
with $m_{s}=0$ and $m_{s}=-1$, $\left|-1,\alpha_{\downarrow}\right\rangle $
is an eigenstate but $\left|0,\alpha_{\uparrow}\right\rangle $ and
$\left|0,\alpha_{\downarrow}\right\rangle $ are not, and therefore
oscillate into one another at an effective nuclear Larmor frequency
$\delta$. The combined action of this nuclear precession, the MW
irradiation and the laser-driven relaxation, results in a redistribution
of the initial populations over the three-level system, biasing the
``bright'' nuclear state population, over its ``dark'' counterpart.
For the example in Fig. \ref{fig:model_scheme}c, a net nuclear magnetization
pointing along the $\left|\alpha_{\downarrow}\right\rangle $ nuclear
quantization axis, defined at weak fields by the hyperfine coupling
to $m_{s}=-1$, is then obtained. Finally, as the MW power is further
increased, a broadband regime where the electron state is flipped
without regards to the nuclear spin state is reached; in such instance
no nuclear polarization enrichment is predicted.

\begin{figure*}
\includegraphics[width=1\textwidth]{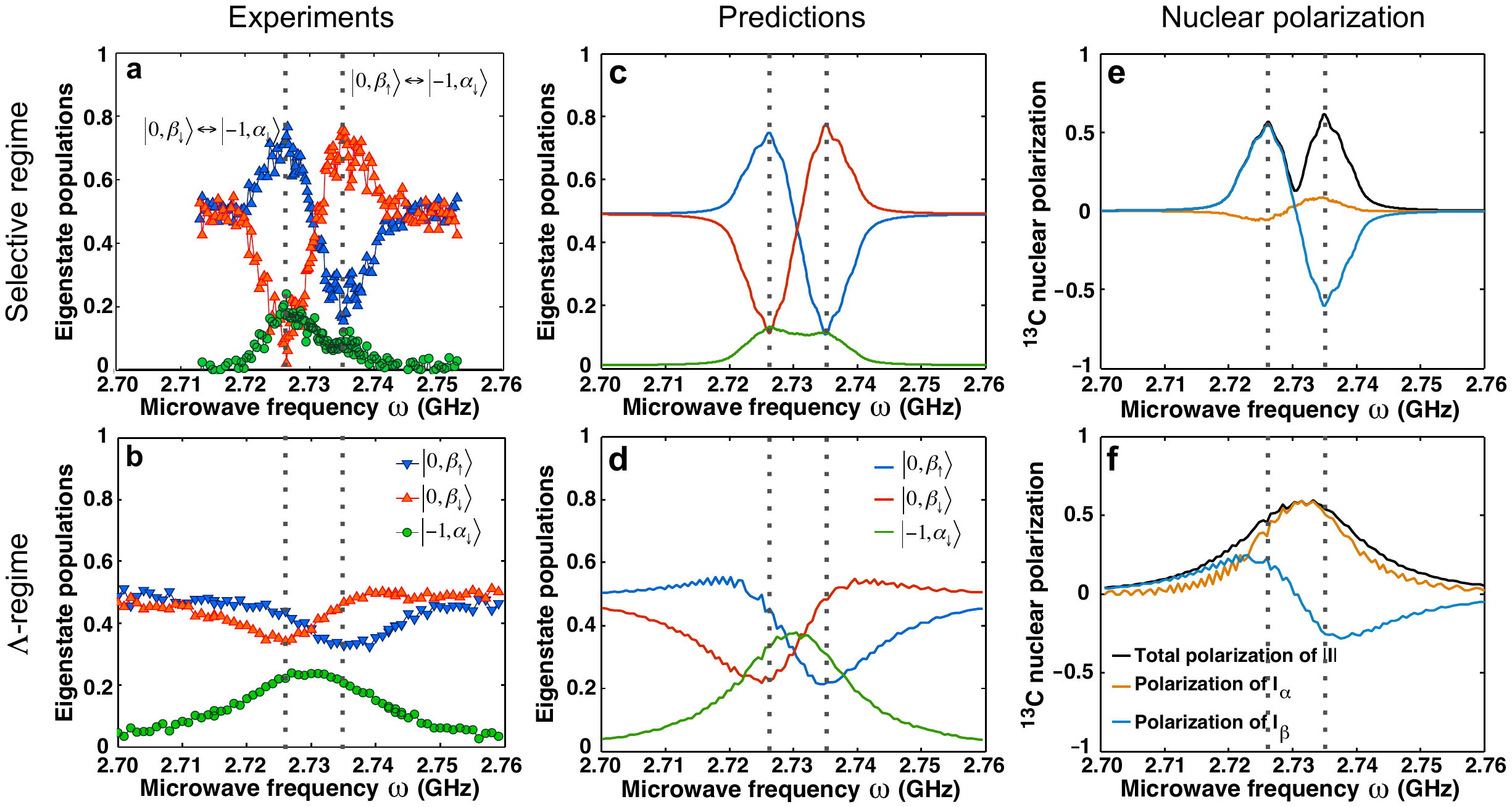}

\protect\caption{\label{fig:ODMR}\textbf{Hyperpolarization of a first-shell nuclear
spin using a MW-driven, optically-pumped NV center.} Experiments were
done by orienting a magnetic field $\left|B_{0}\right|=4.04$mT at
$\theta\approx42^{\circ}$ and $\phi\approx85^{\circ}$ polar and
azimuthal angles with respect to the zero-field tensor $\mathbf{D_{0}}$
of the NV center. The effective $^{13}$C Larmor is then $\delta\approx8.8$
MHz, while the first-shell hyperfine interaction is $\mbox{\ensuremath{\Delta}}\approx130$
MHz. Two MW power strengths were calibrated corresponding to Rabi
frequencies of $\Omega\approx1.4\,\mbox{MHz}<8.8\,\mbox{MHz}\approx\delta$
MHz (top panels) and $\Omega\approx11.9\,\mbox{MHz}>8.8\,\mbox{MHz}\approx\delta$
MHz (bottom panels), corresponding to the selective and the $\Lambda$-regime,
respectively. (\textbf{a},\textbf{b}) Optically detected experiments
determining the eigenstate populations at the end of the polarization
phase during 30$\mu$s as a function of the MW frequency $\omega$.
(\textbf{c},\textbf{d}) Predicted eigenstate population spectra calculated
without free parameters, i.e. using calibrated and known couplings,
and a HF tensor was taken from Ref. \cite{Shim2013}. In the selective
regime (top row), the $\left|-1,\alpha_{\downarrow}\right\rangle -$population
spectrum exhibits two peaks corresponding to the $\left|0,\beta_{\uparrow,\downarrow}\right\rangle \leftrightarrow\left|-1,\alpha_{\downarrow}\right\rangle $
transition frequencies, whereas in the $\Lambda-$regime (bottom row)
the $\left|-1,\alpha_{\downarrow}\right\rangle -$population spectrum
has a single peak centered half-way between the $\left|0,\beta_{\uparrow,\downarrow}\right\rangle \leftrightarrow\left|-1,\alpha_{\downarrow}\right\rangle $
transition frequencies. (\textbf{e},\textbf{f}) Derived $^{13}$C
polarization spectra confirming the polarization directions along
the $\left|\beta_{\downarrow,\uparrow}\right\rangle $-state axis
(top) and the $\left|\alpha_{\downarrow}\right\rangle $-state axis
(bottom), for the selective and $\Lambda$ regimes, respectively.}
\end{figure*}

\textbf{Nuclear hyperpolarization in the vicinity of the NV-center.
}To demonstrate these features, the local $^{13}$C polarization achieved
via this HF-mediated polarization transfer was probed by optically
detected magnetic resonance experiments on a single NV center coupled
to a first-shell $^{13}$C nuclear spin, characterized by a strong
HF coupling $\mbox{\ensuremath{\Delta}}\approx130$ MHz \cite{Felton2009,Shim2013}.
$^{13}$C polarization was measured on a crystal lattice rotated on
purpose away from the external magnetic field by arbitrary azimuthal
and polar angles. The electron spin state was initialized to the $m_{s}=0$
state by laser light, and simultaneously irradiated with MWs for 30
$\mu$s. The eigenstate populations were then optically measured as
a function of the MW frequency $\omega$ over a range covering the
various eigenstate transitions frequencies (Fig. \ref{fig:ODMR}).
Two MW powers were used in these experiments, corresponding to the
selective and $\Lambda$-regimes shown in Fig. \ref{fig:model_scheme}.
For the selective regime a $^{13}$C hyperpolarization pointing along
the $\left|\beta_{\uparrow}\right\rangle $ or $\left|\beta_{\downarrow}\right\rangle $
directions defined by the $m_{s}=0$ eigenstate manifold is reached,
depending on the MW irradiation frequency. For the $\Lambda-$regime
nuclear hyperpolarization is also reached but now aligned along an
$\left|\alpha_{\downarrow}\right\rangle $ axis, defined by the nuclear
eigenstates in the $m_{s}=-1$ manifold. The double peak pattern of
the eigenstate population and $^{13}$C polarization spectra is characteristic
of the selective regime. These two $^{13}$C spin alignment maxima
arise when the MW is on resonance with the $\left|0,\beta_{\uparrow,\downarrow}\right\rangle \leftrightarrow\left|-1,\alpha_{\downarrow}\right\rangle $
transition frequencies, leading to opposite nuclear polarization directions.
By contrast, in the $\Lambda-$regime, a single $^{13}$C spin alignment
peak centered between the $\left|0,\beta_{\uparrow,\downarrow}\right\rangle \leftrightarrow\left|-1,\alpha_{\downarrow}\right\rangle $
transition frequencies is observed, evidencing that these two transitions
are affected by the MW, and leading to a nuclear spin alignment defined
by the $\left|-1,\alpha_{\downarrow}\right\rangle $ nuclear quantization
axis. 

\begin{figure*}
\includegraphics[width=0.8\textwidth]{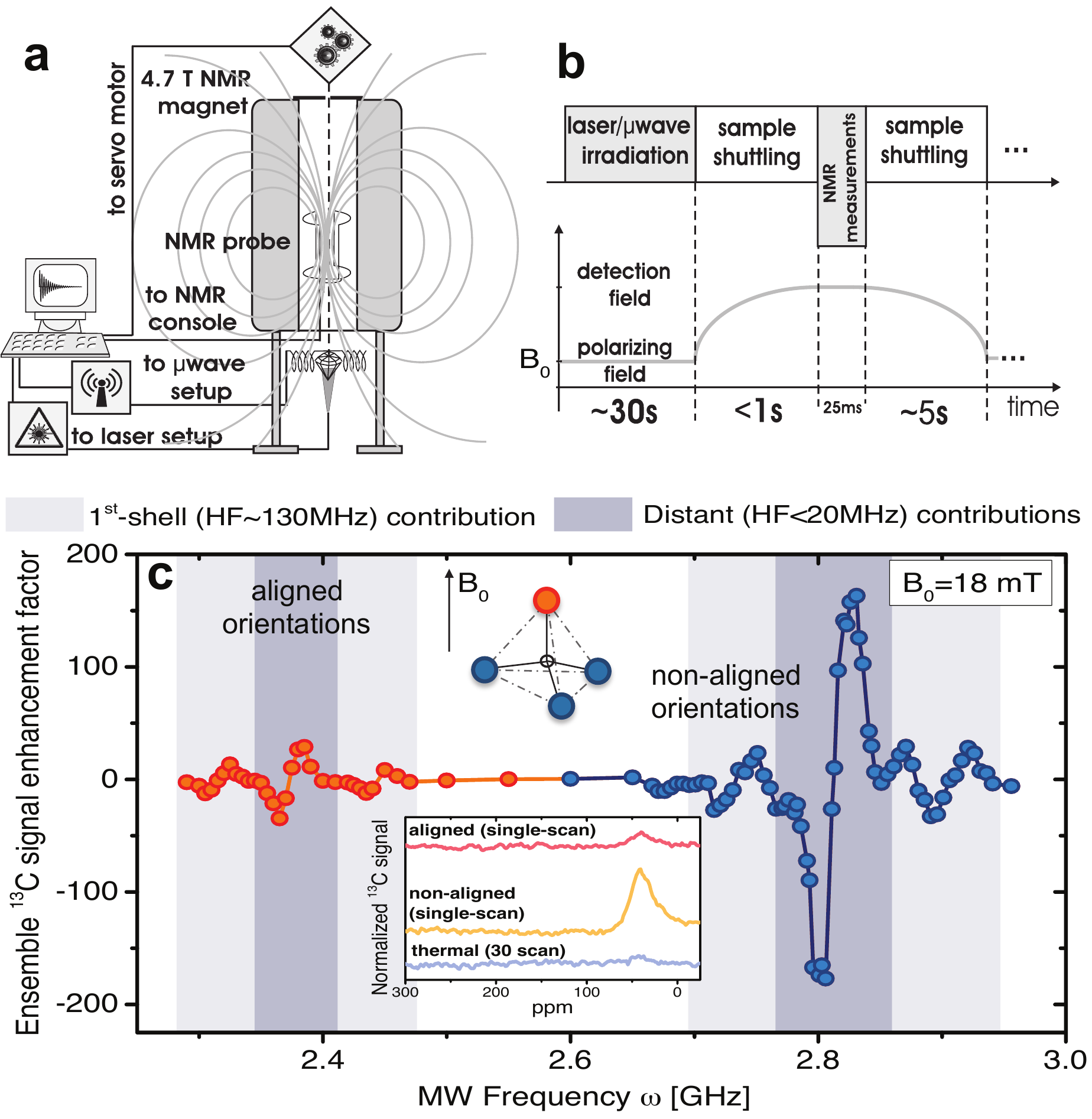}

\protect\caption{\label{fig:13cvsMWfreq}\textbf{Acquiring ensemble $^{13}$C polarization
spectra for varying NV orientations with respect to $B_{0}$.} (\textbf{a})
Opto-NMR setup and (\textbf{b}) detection sequence used in these experiments.
During the polarization transfer phase, the entire single-crystal
diamond is irradiated with laser light and MW underneath the NMR magnet
at a low $B_{0}$. The hyperpolarized diamond is then shuttled (in
$<1$ sec) into a 4.7 T superconducting magnet, to directly detect
its macroscopic $^{13}$C magnetization via a spin-echo sequence.
The low $B_{0}$ magnetic field is aligned to one of the NV-center
orientations (in red), while the other three orientations (in blue)
subtend an angle of $\approx109^{\circ}$ with respect to the field.
(\textbf{c}) Typical $^{13}$C polarization enhancement patterns observed
by NMR as a function of the MW frequency $\omega$ with signals normalized
with respect to the thermal $^{13}$C response at 4.7 T (inset). The
left part of the plot corresponds the nuclear polarization generated
by $\left|m_{S}=0\right\rangle \rightarrow\left|m_{S}=-1\right\rangle $
MW transitions for the aligned orientation (red circles), while the
right part corresponds to nuclear polarization enhanced via the three
non-aligned, equivalent orientations (blue circles). The $\approx$1:3
intensities ratio reflects the relative abundances of aligned and
non-aligned sites in the diamond's tetrahedral structure. In each
of the patterns the ``central'' peaks represent bulk nuclear hyperpolarization
pumped via $^{13}$C spins coupled with HF interactions lower than
20 MHz, while the ``outer'' peaks originate from first-shell $^{13}$C's
whose HF splitting is $\approx130$ MHz \cite{Felton2009,Shim2013}.
The ``anti-phase'' structure of each of these peaks corresponds to
the MW transitions $\left|0,\beta_{\downarrow}\right\rangle \leftrightarrow\left|-1,\alpha_{\downarrow}\right\rangle $
and $\left|0,\beta_{\uparrow}\right\rangle \leftrightarrow\left|-1,\alpha_{\downarrow}\right\rangle $
at one side of the central peaks, and to the $\left|-1,\alpha_{\uparrow}\right\rangle $
state at the other side. The inset shows NMR spectra obtained for
a thermally polarized sample, and at the maxima of the ``central''
peaks for the aligned and non-aligned orientations.}
\end{figure*}

\begin{figure}
\includegraphics[bb=0bp 0bp 466bp 489bp,width=1\columnwidth]{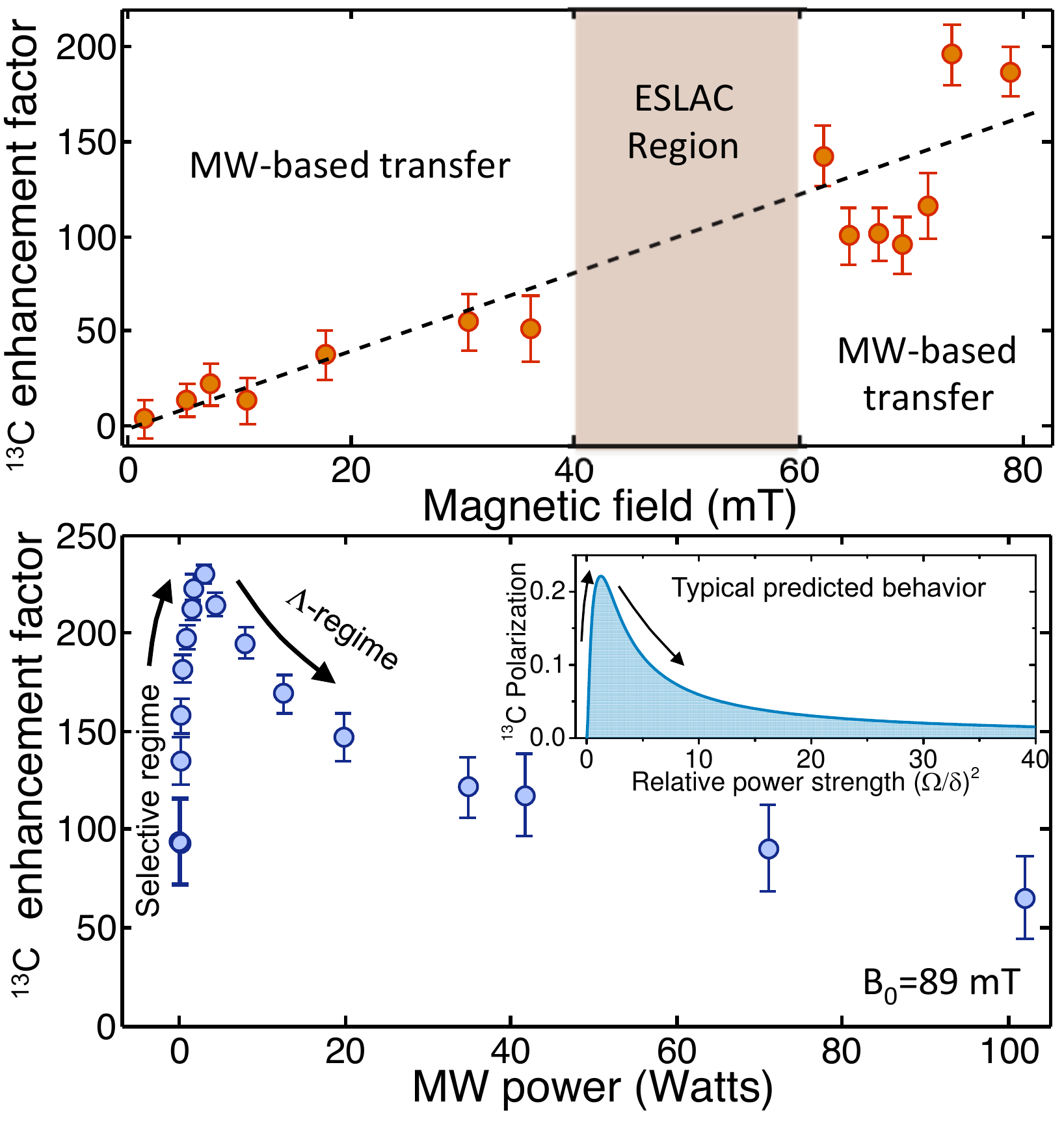}

\protect\caption{\label{fig:13Cvspower}\textbf{Ensemble $^{13}$C-magnetization for
different magnetic field strengths and microwave powers.} The experimental
points of these panels correspond to the maximum signal of the central
peak in the aligned orientation shown in Fig. \ref{fig:13cvsMWfreq}c.
(\textbf{a}) $^{13}$C polarization as a function of the magnetic
field strength (normalized to its thermal 4.7 T counterpart and measured
up to our maximum operating field, corresponding to $\approx$100
mT). The dashed line is a guide to the eye. Within the shaded\textbf{
}ESLAC region \cite{Dutt2007,Jacques2009,Fischer2013} polarization
transfer was observed, but such polarization transfer scheme derives
from level anti-crossing and is not considered by the present model.
(\textbf{b}) $^{13}$C hyperpolarization as a function of the MW power
at a fixed field of $\approx89$ mT. The inset shows the typical predicted
behavior by the model, where the $^{13}$C polarization grows with
MW power within the selective regime, but decreases as the $\Lambda$-regime
is reached.}
\end{figure}

\textbf{Bulk nuclear hyperpolarization at variable fields and orientations.
}The polarized NV-center is embedded in a network of dipole-dipole
interacting nuclear spins, allowing an eventual propagation of the
hyperpolarization throughout the nuclear bulk ensemble. Achieving
an effective $^{13}$C polarization enhancement throughout an entire
diamond could have important practical NMR and MRI consequences. Consequently
we investigated whether distant $^{13}$C spins can benefit from the
polarization transfer mechanisms observed in Fig. \ref{fig:ODMR}
at a local level. The bulk macroscopic $^{13}$C magnetization was
directly measured by mechanically shuttling the diamond from low fields,
where MW and laser fields were applied to accomplish the electron\textbf{\textrightarrow }nuclear
polarization, into a 4.7 T magnet enabling bulk $^{13}$C NMR detection
(Fig. \ref{fig:13cvsMWfreq}a-b). 
Figure \ref{fig:13cvsMWfreq}c presents results arising from a typical
experiment, showing the dependence of the ensemble $^{13}$C-magnetization
on the MW frequency $\omega$. For consistency, measurements were
carried out with the polarizing $B_{0}$ field aligned along one of
the NV center orientations, meaning that the remaining three orientations
of the crystal form an angle of $\approx109^{\circ}$ away from the
field direction. The nuclear polarization spectrum (Fig. \ref{fig:13cvsMWfreq}c)
shows two regions of enhancement, corresponding to the aligned and
non-aligned orientations of the NV defects. Both regions show similar
patterns of multiple positive and negative peaks, corresponding to
different HF couplings and MW-induced transitions. Both $^{13}$C
patterns contain major ``central'' peaks, and minor ``outer'' peaks
detuned by $\approx\pm60$ MHz from their centers. The central peaks
correspond to bulk $^{13}$C nuclei being hyperpolarized via $\lesssim20$
MHz HF interactions, i.e. via $^{13}$C spins positioned at or beyond
the NV's 2nd-shell \cite{Smeltzer2011,Dreau2012}. The outer peaks
also correspond to bulk $^{13}$C polarization, but arriving this
time via first-shell $^{13}$C's whose HF splitting is $\approx130$
MHz \cite{Felton2009,Shim2013}. In this instance the bulk $^{13}$C
magnetization exhibits a sign-flip pattern characteristic of the ``selective''
regime akin to the local $^{13}$C polarization pattern shown in Fig.
\ref{fig:ODMR}c -although these are ``bulk'' $^{13}$C resonances
detected on the whole diamond, and not ``single spin'' measurements.
This demonstrates that $^{13}$C ensemble hyperpolarization can be
derived via nuclear-nuclear dipole couplings involving distant spins
to the electron, but also via first-shell $^{13}$C spins. The latter
is a remarkable fact, given that in this case the build-up of bulk
polarization needs to overcome >60 MHz detuning frequencies vis-a-vis
the polarization-transfer ``bridgehead''. The timescales that are
then needed to achieve a bulk polarization build-up, about 30 sec,
are orders of magnitude longer than the first-shell electron\textrightarrow $^{13}$C
polarization transfer time ($\sim$ ms). This difference in timescales
explains how a low-concentration, highly-detuned species like the
first-shell $^{13}$C, suffices to polarize the bulk sample examined
by the NMR experiments. Figure \ref{fig:13Cvspower} further demonstrates
the versatility of the method to yield bulk nuclear hyperpolarization,
as evidenced by the achievement of enhanced $^{13}$C NMR signals
over a broad range of polarizing magnetic fields and of MW powers.
Notice that the nuclear polarization reveals a systematic increase
as a function of $B_{0}$, as well as a clear optimum for the MW power
(Fig. \ref{fig:13Cvspower}b) consistent with a transition from a
selective regime in which the polarization grows with MW power, to
a $\Lambda-$regime in which it decreases with the MW power (see inset).

\section*{Discussions}

A versatile method for hyperpolarizing nuclear spin ensembles at room
temperature using NV centers in diamond, is proposed and demonstrated.
The method relies on the combined action of continuous laser-light
and microwave irradiation and exploits the asymmetries of the hyperfine
interaction imparted by the $m_{s}=0$ and $m_{s}=\pm1$ electronic
spin states. Nuclear spin hyperpolarization is obtained for a broad
range of magnetic field strengths and arbitrary field orientations
with respect to the diamond lattice, as well as for a wide range of
hyperfine interactions. These principles may facilitate a number of
applications, particularly for ensemble and nanoscale NMR/MRI \cite{Mamin2013,Staudacher2013,Loretz2014,Abrams2014,Loretz2014a,Sushkov2014},
and for dissipative nuclear-state preparation of ensemble quantum
memories for quantum information processing \cite{Verstraete2009,Lin2013}.
They might also find applicability in other kinds of cross-polarization
spin scenarios. Besides demonstrating the method's capability at a
local level, it was shown that nuclear hyperpolarization can be extended
throughout the bulk ensemble by spin-diffusion -even when targeting
1$\mbox{\ensuremath{^{st}}}$ shell nuclear spins strongly shifted
by HF couplings. This provides a new tool for understanding the complex
non-equilibrium dynamics of many-body systems, where a local polarization
can be created in a controlled way and its spreading into the bulk
be monitored \cite{Richerme2014,Jurcevic2014}.

\section*{Methods\medskip{}
}

\textbf{Single crystal sample.} For the single NV detection experiments
the sample used was a commercially available, untreated, type IIa,
electronic-grade natural abundance $^{13}$C diamond crystal,(dimension:
2$\times$2$\times$0.3 mm$^{3}$). For the NMR experiments, an isotopically
enriched (10\% $^{13}$C) type Ib high-pressure, high-temperature
(HPHT) diamonds was used. The enriched crystal was grown by the temperature
gradient method at a pressure of 6 GPa and a temperature 1700 K from
Ni-2 wt.\% Ti solvent using a powdered mixture of $^{13}$C-enriched
and natural abundance graphic carbon. These samples were irradiated
at room temperature (2 MeV, 10 h, total fluence $8\times10^{17}$
e/cm$^{2}$) and annealed (2 hours, 1000 $^{\circ}$C). Subsequent
examination by confocal laser microscopy confirmed that the two most
abundant paramagnetic impurities consisted of electrically neutral
single substitutional nitrogen atoms (P$_{1}$), and charged nitrogen
atoms next to a lattice vacancy forming an optically active color
center ($NV^{-}$). The concentrations of these impurities were both
$<$ 5 ppm.

\medskip{}

\textbf{System Hamiltonian. }At room temperature the NV-center energy
level structure exhibits an electronic triplet as the ground state
($^{3}A_{2}$) \cite{Dutt2007,Jacques2009,Fischer2013}. The quantum
Hamiltonian of a single NV defect ($\hat{S}$) and one $^{13}$C nucleus
($\hat{I}$) can thus be described as $\hat{H}=D_{0}\:\hat{S}_{z}^{2}+\gamma_{e}\:\vec{S}\cdot\vec{B}+\gamma_{e}\:\vec{I}\cdot\vec{B}+\vec{I}\cdot\mathbf{A}\cdot\vec{S}$.
Here $D_{0}=2.87$ GHz is the zero-field splitting term, $\gamma_{e}$
and $\gamma_{n}$ the electronic and nuclear gyromagnetic ratios,
$\vec{B}$ is the magnetic field vector, and $\mathbf{A}$ a hyperfine
tensor that depends on the specific NV and nearby $^{13}$C spin.
For simplicity, we consider first a magnetic field $\vec{B}$ aligned
with the axis of the zero-field tensor $D_{0}$ and the secular approximation
$\left|D_{0}\pm\gamma_{e}B_{0}\right|\gg A_{uv}$. This Hamiltonian
simplifies to $H=D_{0}S_{z}^{2}+\left(\gamma_{e}S_{z}+\gamma_{n}I_{z}\right)B_{0}+A_{zz}S_{z}I_{z}+A_{zx}S_{z}I_{x}$,
where the axis $x$ was chosen such that $A_{zy}=0$. The feature
enabling the magnetization transfers illustrated in Fig. \ref{fig:model_scheme},
is that in the $m_{s}=0$ manifold the eigenstates have a nuclear
component $\left|\beta_{\uparrow}\right\rangle =\left|\uparrow\right\rangle $
and $\left|\beta_{\downarrow}\right\rangle =\left|\downarrow\right\rangle $
exhibiting a nuclear Zeeman splitting $\delta=\gamma_{n}B_{0}$; whereas
in the $m_{s}=-1$ manifold the $\left|\alpha_{\uparrow}\right\rangle $
and $\left|\alpha_{\downarrow}\right\rangle $ nuclear components
of the eigenstates are quantized on a different axis determined by
the hyperfine tensor. For the $\left|\beta_{\uparrow,\downarrow}\right\rangle $
states the eigenstate splitting $\delta$ may also depend on non-secular
corrections of the HF coupling \cite{Shim2013}, which may also lead
to an effective tilt of the nuclear spin quantization away from the
magnetic field $\vec{B}$. By contrast, the $\left|\alpha_{\uparrow,\downarrow}\right\rangle $
states are not significantly modified by the non-secular HF terms
if $\left|D_{0}\pm\gamma_{e}B_{0}\right|\gg A_{uv}$, i.e. on a condition
far from the level anti-crossings \cite{Jacques2009,Fischer2013,Wang2013}
(we do not consider here the more complicated scenario when the level
anti-crossings conditions are fulfilled). If the magnetic field is
not aligned with the $\mathbf{D_{0}}$ tensor, the level structure
discussed in Fig. \ref{fig:model_scheme} remains even if the definitions
of $\left|\beta_{\uparrow,\downarrow}\right\rangle $ and $\left|\alpha_{\uparrow,\downarrow}\right\rangle $
will change; the crucial characteristic is that they will still have
different quantization axes and different eigenstate splittings $\delta$
and $\Delta$. \medskip{}

\textbf{Single NV optical setup.} The single NV detection was conducted
on a home-built confocal microscope. Continuous wave green laser light
($\lambda$ = 532 nm) was switched by an acousto-optic modulator (Isomet,
rise/fall times $\approx$30 ns) coupled into a single mode optical
fiber and focused onto the diamond sample by an oil-immersion microscope
objective (Olympus, NA=1.35). The objective was placed on a three-axis
piezo stage (Npoint) enabling a scan range of the $x,\, y,\, z$ axes
of 250$\times$250$\times$25$\mu$m with a resolution of $\sim$1
nm. The emerging red fluorescence signal was focused into a single-photon
counting module (Perkin-Elmer, dark count rate 100 counts/sec). To
induce transitions within the ground state triplet during the polarization
and detection sequences, MW fields were produced by two copper wires
attached to the diamond (diameter $\sim$ 50 $\mu$m). During the
hyperpolarization process, the laser intensity was reduced to about
5\% of its saturation intensity ($P_{sat}\sim$1mW at the objective's
back), in order to avoid light-induced nuclear-state depopulation.\medskip{}

\textbf{Eigenstate populations and nuclear polarization determination
of the single NV-center experiments}. The eigenstates\textquoteright{}
populations were obtained by performing a series of complementary
pulse sequences involving selective MW irradiation on eigenstate-specific
transition frequencies, followed by an optical readout of the NV center
populations. With these selective experiments and a linear system
of equations, the populations of the three active eigenstates ($\left|0,\beta_{\uparrow,\downarrow}\right\rangle $
and $\left|-1,\alpha_{\downarrow}\right\rangle $) could be reconstructed.
To perform these calculations, the fluorescence levels at the end
of the different polarization sequences were measured, and properly
normalized by executing the schemes with and without MW irradiation.
In the selective regime, the MW Rabi frequency $\Omega=1.4$ MHz was
lower than the HF splitting with the $^{14}$N nuclear spin (2.16
MHz). Therefore, to address the latter\textquoteright s three nuclear
spin orientations, the MW source was further modulated to create the
three corresponding frequencies spaced by 2.16 MHz. The wiggles in
Fig. \ref{fig:ODMR}a,c,d are consequence of this irradiation mechanism.
The magnetic field strength and orientation were optimized to be large
enough for allowing selective excitation independent of the host $^{14}$N
nuclear spin ($\delta>4-5$ MHz), and small enough to allow mutual
excitation (i.e $\Lambda-$regime) with reasonable MW power ($\delta<10$
MHz). The static and MW magnetic fields and the laser-pumping rate
were calibrated from ODMR spectra, and from time-resolved measurements
of the Rabi oscillations and the laser-pumping evolution. The values
of the static and MW fields are given in the caption to Fig. \ref{fig:ODMR},
while the optical pumping rate was 1/(3$\mu$s) and had a pumping
efficiency of $\approx90$\%. The curves in Fig. \ref{fig:ODMR}c-f
were obtained by solving the spins\textquoteright{} quantum evolution
based on a quantum master equation involving the total 6-level model
of a single NV center coupled to a single $^{13}$C nuclear spin.
The bare system Hamiltonian used was introduced in the Methods section;
simulations were performed on the basis of this Hamiltonian in the
rotating frame of the MW irradiation, with frequency parameters as
discussed in the main text. The pumping process was modeled as a non-Hermitian
relaxation process using a Lindblad form reflecting the incoherent
transition from the $m_{s}=\pm1$ states to the $m_{s}=0$ state,
using the experimentally determined pumping rate. The HF tensor of
the first-shell $^{13}$C spin was taken from Ref. \cite{Shim2013}.
For the selective regime, the $^{14}$N HF splitting was considered
phenomenologically as an ensemble superposition of the signal coming
from the three frequencies used in the combed MW irradiation.\medskip{}

\textbf{Setup for the ensemble-detected nuclear spin polarization
experiments. }In a single-crystal diamond, NV centers can be separated
into sub-ensembles corresponding to four magnetically inequivalent
orientations. The diamond sample used for the optically pumped NMR
measurements had a polished {[}100{]} surface orientation. The sample
was rotated in $B_{0}$ until one of the NV center orientations was
aligned to the magnetic field, while the three remaining sub-ensembles
were degenerate at an angle of $\sim$ 109$^{\circ}$ forming the
``non-aligned'' orientations. The weak magnetic field in these experiments
($B_{0}$ = 0-100 mT) was provided by the stray field of the superconducting
magnet used to perform the bulk high-field NMR measurements, connected
by a compensating low-field coil. The laser light was produced by
a diode pumped solid-state laser (Verdi-V10, $\lambda$=532 nm) coupled
into an optic fiber and then focused to either partially or fully
illuminate the diamond crystals. To avoid excessive heating during
the optical pumping the sample was immersed into a transparent water-filled
flask and vented with pressurized air to keep the temperature constant.
For the application of continuous wave MW irradiation, radio- and
MW signals from a broadband frequency synthesizer were amplified by
up to +30 dB for frequencies $\omega>$ 600 MHz and up to +43 dB for
$\omega<$ 600 MHz. These amplified signals were then fed into a Helmholtz
coil electronic circuit tuned to 50 $\Omega$. 

The bulk $^{13}$C experiments began with a combined laser and MW
irradiation ($10-30$ s) to optically initialize the electronic spins
and induce a transfer of the ensuing hyperpolarization from the electronic
to the nuclear spins. The sample was then rapidly (<0.5 s) shuttled
from the polarizing field $B$ to the ``sweet spot'' of the high-field
(4.7 T) NMR magnet, where the nuclear spins were subject to a pulsed
detection using an NMR probe with a Helmholtz coil configuration tuned
to 50.55 MHz. The observed NMR signal consists of a single optically-enhanced
$^{13}$C spectral line resonating at ca. 60 ppm with a linewidth
between 1 (natural abundance) and 3 kHz (10\% enriched crystal). The
sample then returned to the polarizing field for a repeated cycle
of optical pumping and MW irradiation for the purpose of signal averaging.\medskip{}

\section*{Acknowledgments}
\begin{acknowledgments}
We thank Shimon Vega, Gershon Kurizki and Analia Zwick for stimulating
discussions. We thank Fedor Jelezko and his group for their guiding
in constructing the single NV-centers confocal microscope, and for
obtaining the typeIIa sample. This work was supported by the DIP Project
710907, ERC Advanced Grant \# 246754, the Kimmel Institute for Magnetic
Resonance, IMOD, the Israel Science Foundation Grant 1142/13, the
generosity of the Perlman Family Foundation, JST and JSPS KAKENHI
(No.26246001). G.A.A. acknowledges the support of the European Commission
under the Marie Curie Intra-European Fellowship for career Development
Grant No. PIEF-GA-2012-328605.\medskip{}

\end{acknowledgments}

\section*{Author contributions}

G.A.A., R.F., and L.F. conceived the polarization transfer scheme.
G.A.A., C.O.B.., R.F., P.L. and L.F. designed the experimental approaches.
P.L. and R.F. implemented and performed the single spin experiments
and analyzed their data. C.O.B. and R.F. implemented the hardware-setup
for the bulk-NMR experiments, and G.A.A. and C.O.B. performed them,
including data processing. G.A.A. and R.F. performed the numerical
simulations. G.A.A., C.O.B., R.F., P.L. and L.F. analyzed and interpreted
the data. H.K, S.O. and J.I. prepared the enriched diamond sample.
G.A.A., C.O.B., R.F. and L.F. wrote the manuscript. All authors comment
on the manuscript and contributed to discussions of its results.

\bibliographystyle{naturemag}
\bibliography{bibliography}

\end{document}